\documentclass[12pt,preprint2]{aastex}
\usepackage{emulateapj5,apjfonts,onecolfloat5}

\slugcomment{Accepted for publication in the Astrophysical Journal
Letters}

\shorttitle{Chandra detection of pulsar J1119--6127 in
G292.2--0.5}

\shortauthors{Gonzalez and Safi-Harb}

\def\chan{$\it{Chandra}$}
\def\g292{G292.2--0.5}
\def\j1119{J1119--6127}
\def\axj{AX J1119.1--6128.5}
\def\j1846{J1846--0258}
\def\b1509{B1509--58}

\begin{document}
\twocolumn[

\title{Chandra detection of the X-ray counterpart of the high magnetic field radio pulsar J1119-6127 in
the supernova remnant G292.2-0.5}
\author{Marjorie Gonzalez\altaffilmark{1,2} and Samar Safi-Harb\altaffilmark{1,3}}
\affil{$^{1}$ Physics and Astronomy Department, University of
Manitoba, Winnipeg, MB, R3T 2N2, Canada; }
\affil{$^{2}$umgonza4@cc.umanitoba.ca (NSERC PGS A fellow);
$^{3}$safiharb@cc.umanitoba.ca (NSERC UFA fellow)}

\begin{abstract}
We report the \chan\ Advanced CCD Imaging Spectrometer detection
of the X-ray counterpart of the high magnetic field, $\sim$1600
year-old, 407 ms radio pulsar J1119--6127 associated with the
supernova remnant \g292. The powerful imaging capability of
\chan\ also unveiled, for the first time, a faint
$3''$$\times$$6''$ pulsar wind nebula (PWN) at energies above
$\sim$1.2 keV. The X-ray emission from the pulsar and its
associated nebula is well described by an absorbed power law model
with a photon index $\Gamma$ = 2.2$^{+0.6}_{-0.3}$. The
corresponding 0.5--10 keV unabsorbed X-ray luminosity is
(5.5$^{+10}_{-3.3}$)$\times$10$^{32}$~ergs~s$^{-1}$ (at 6~kpc).
When compared to two other pulsars with similar spin and magnetic
properties, J1119--6127 stands out as being the least efficient at
turning rotational kinetic energy into X-ray emission. This study
shows that high magnetic field radio pulsars can be significant
X-ray emitters and \chan\ is needed to study the emission
properties of the pulsars and associated faint PWNe.

\end{abstract}



]

\section{Introduction}
For over three decades the Crab has been viewed as the paradigm
for young pulsars: a fast rotating neutron star with surface
dipole magnetic field strength of 10$^{12}$~G, injecting a
relativistic magnetized wind of particles into its surroundings.
The interaction of this wind with the surrounding medium creates a
synchrotron nebula, referred to as a pulsar wind nebula (PWN).
PWNe provide a unique laboratory that probes the properties of
their powering engines, the physics of relativistic pulsar winds
and their interaction with the interstellar medium (see e.g.
Safi-Harb 2002 for a review). Recent observations have shown that
a good fraction of young pulsars exhibits properties unlike the
Crab. In spite of their youth, they have much larger spin periods
and dipole magnetic fields (see e.g. Camilo et al. 2000). The
search for their X-ray counterparts and any PWNe associated with
them sheds light on their high-energy properties and the way these
pulsars deposit their energies into their surroundings.

The radio pulsar (PSR) J1119--6127 was discovered in the Parkes
multi-beam pulsar survey (Camilo et al. 2000). It has a rotation
period of 407ms, characteristic age of 1600 yrs, surface magnetic
field strength of 4.1$\times$10$^{13}$~G, and spin-down luminosity
$\dot{E}$ of 2.3$\times$10$^{36}$ erg~s$^{-1}$. This is an
interesting and unusual object: although it is extremely young, it
displays a relatively large period and magnetic field compared to
Crab-like pulsars. Furthermore, no radio emission from a PWN was
found in spite of the search conducted with the Australia
Telescope Compact Array (ATCA). Its observed upper limit is below
what might at first be expected from the pulsar's characteristics
(Crawford et al. 2000). The ATCA observation also identified a
non-thermal shell of $15'$ in diameter surrounding the pulsar
classified as a new supernova remnant (SNR), \g292. Previous X-ray
observations, performed with the $\it{ASCA}$ and $\it{ROSAT}$
satellites (Pivovaroff et al. 2001), showed extended emission
associated with the radio SNR, as well as a hard point-like
$\it{ASCA}$ source offset $\sim$$1\farcm5$ from the radio pulsar.
The large offset made their association uncertain.

In this paper, we report the detection of the X-ray counterpart of
the radio pulsar and show the first evidence for a PWN associated
with it. In \S2, we describe the observation and data analysis. In
\S3, we derive the distance using our spectral fits to the X-ray
source and SNR interior, then discuss the X-ray properties of this
system in comparison with other pulsars with similar spin and
magnetic field properties. We also show that the previously
reported $\it{ASCA}$ source is most likely the unresolved
counterpart of the radio pulsar, contaminated by emission from its
surroundings.

\begin{deluxetable}{lc}
\tablewidth{0pt} \tablecaption{Power-law spectral model fit for
the X-ray counterpart of PSR J1119--6127\tablenotemark{a}}
\tablehead{ \colhead{Parameter} &\colhead{Value ($\pm$ 90\%)}}
\startdata
$N_{H}$ (10$^{21} $cm$^{-2}$)       &   9$^{+5}_{-3}$\\
Photon index, $\Gamma$              &   2.2$^{+0.6}_{-0.3}$\\
Norm$_{1 keV}$ (photons keV$^{-1}$ cm$^{-2}$ s$^{-1}$)  & (3.1$^{+3.3}_{-1.5}$)$\times$10$^{-5}$\\
$f_{abs}$ (0.5--10.0 keV, ergs~cm$^{-2}$~s$^{-1}$)  & (6.6$^{+13.4}_{-4.8}$)$\times$10$^{-14}$ \\
$f_{unabs}$ (0.5--10.0 keV, ergs~cm$^{-2}$~s$^{-1}$)    & (1.3$^{+2.4}_{-0.8}$)$\times$10$^{-13}$ \\
$\chi^{2}_{\nu}$(dof)                   & 0.57 (16)
\enddata
\tablenotetext{a}{Includes point source and extended component,
see \S2.1 for details.}
\end{deluxetable}

\section{Observation and Data Analysis}
The field around pulsar J1119--6127 was observed with \chan\ on
2002 March 31--April 1. The coordinates of the X-ray source
detected with $\it{ASCA}$ were positioned at the aimpoint of the
back-illuminated S3 chip of the Advanced CCD Imaging Spectrometer
(ACIS). The CCD temperature was --120$\degr$C with a frame readout
time of 3.2 sec in ``very faint'' mode. We applied the
CTICORRECTIT tool to the original event 1 raw data (Townsley et
al. 2000) in order to correct for charge transfer inefficiency.
The data were then calibrated using standard CIAO 2.2 routines.
The resulting effective exposure time was 47~ks.

Figure 1 shows the $\it{ASCA}$ (left) and \chan\ (right) images of
\g292. The 2.0--10.0 keV $\it{ASCA}$ image has been smoothed using
a Gaussian with $\sigma$ = $45''$\footnote{See Pivovaroff et al.
(2001) for a detailed analysis of the $\it{ASCA}$ data}. The
\chan\ image was obtained as follows: the data were divided into
individual images in the soft (0.5--1.15 keV, red), medium
(1.15--2.3 keV, green), and hard (2.3--10.0 keV, blue) bands. Each
image was adaptively smoothed using a Gaussian with $\sigma$=$1''$
for significance of detection $>$5 and up to $\sigma$=$10''$ for
significance down to 3. A broadband (0.5--10.0 keV) background
image was produced using the blank-sky datasets available in CALDB
v2.12. The resulting background image was divided into the same
energy bands and then subtracted from its corresponding (source +
background) image. The individual background-subtracted images
were finally combined to produce the image shown in Figure 1. This
image shows several resolved X-ray sources surrounded by diffuse
emission from the interior of the remnant.

\subsection{PSR J1119--6127}
In Figure 2 (left) we show a close-up image of the ACIS field
around the radio pulsar J1119--6127. We find an X-ray source at
$\alpha$$_{J2000}$ = 11$^{h}$19$^{m}$14$\fs$4 and
$\delta$$_{J2000}$ = --61$\degr$$27'$49$\farcs$7, with a 90\%
error radius of 1$\farcs$3. This source is the brightest one in
the field and it lies 0$\farcs$7 away from the coordinates of the
radio pulsar ($\alpha$$_{J2000}$ = 11$^{h}$19$^{m}$14$\fs$30 and
$\delta$$_{J2000}$ = --61$\degr$$27'$49$\farcs$5, 0$\farcs$3
error, Camilo et al. 2000). From the positional coincidence of the
X-ray source with the radio pulsar, the low probability of a
chance alignment ($\sim$1$\times$10$^{-4}$), the evidence of a PWN
associated with it and the nature of its spectrum (see below), we
believe this source to be the X-ray counterpart of PSR
J1119--6127.

Within a radius of $3\farcs4$ centered at the X-ray coordinates
and including the point source and extended component, we obtained
285 background subtracted events in the 0.5--10.0 keV band
translating to a count rate of (6.0 $\pm$
0.4)$\times$10$^{-3}$~cts~s$^{-1}$. Pile-up effects are then
negligible.  We fit the spectrum with
XSPEC\footnote{http://heasarc.gsfc.nasa.gov/docs/xanadu/xspec/index.html}
in the 0.5--10.0 keV band, using a minimum of 15 counts per bin.
An absorbed power law model provided a better fit than thermal
models. The best-fit spectral parameters are summarized in Table 1
and the spectrum is shown in Figure 2 (right). The corresponding
unabsorbed fluxes are
(1.4$^{+2.6}_{-0.8}$)$\times$10$^{-13}$~ergs~cm$^{-2}$~s$^{-1}$
(0.2--2.4 keV) and
(1.3$^{+2.4}_{-0.8}$)$\times$10$^{-13}$~ergs~cm$^{-2}$~s$^{-1}$
(0.5--10.0 keV). All errors throughout the paper are at the 90\%
confidence level unless otherwise specified. We note that by
examining Figure 2 (right), we find residuals at $\sim$1.4 keV
indicating that the spectrum could be partly thermal. However,
these residuals become insignificant when extracting a spectrum
from a smaller region. The poor statistics did not allow us to fit
multi-component models. A detailed investigation of this emission
will have to await a deeper exposure.

\begin{deluxetable}{lccc}
\tablewidth{0pt} \tablecaption{Observed parameters for PSRs
J1119--6127, J1846--0258 and B1509--58} \tablehead{
\colhead{Parameter} & \colhead{J1119--6127} &
\colhead{J1846--0258} & \colhead{B1509--58}} \startdata
Spin period, $P$ (ms)                   &  408  &  324  &  150  \\
Period derivative, $\dot{P}$        & 4$\times$10$^{-12}$ & 7.1$\times$10$^{-12}$ & 1.5$\times$10$^{-12}$ \\
Surface magnetic field, $B$ (G)     & 4.1$\times$10$^{13}$ & 4.8$\times$10$^{13}$  & 1.5$\times$10$^{13}$ \\
Characteristic age, $\tau_{c}$ (yr)     & 1600  &   980--1700 & 1700  \\
Spin-down luminosity, $\dot{E}$ (erg~s$^{-1}$)  & 2.3$\times$10$^{36}$  & 7.9$\times$10$^{36}$  & 18$\times$10$^{36}$   \\
Braking index, $n$                      & 2.91  & 1.86--2.48    & 2.8   \\
Distance & $\sim$6~kpc & $\sim$19~kpc & $\sim$5~kpc \\
$N_{H}$ (10$^{22}$ cm$^{-2}$)           & $\sim$0.9 & $\sim$4   & $\sim$1   \\
Photon index    & $\Gamma_{psr}$$\gtrsim$2.2, $\Gamma_{pwn}$$\lesssim$2.2 & $\Gamma_{psr}$$\sim$1.4, $\Gamma_{pwn}$$\sim$1.9  & $\Gamma_{psr}$$\sim$1.4(?), $\Gamma_{pwn}$$\sim$2.05 \\
Radio/X-ray PWN?  & no / yes     & yes / yes    &    yes / yes    \\
0.5--10.0 keV efficiency, $\epsilon$ ($L_{X}$/$\dot{E}$)    & $\epsilon_{psr+pwn}$ $\lesssim$ 0.001 &  $\epsilon_{psr}$$\sim$0.016, $\epsilon_{pwn}$$\sim$0.065    & $\epsilon_{psr}$$>$0.001, $\epsilon_{pwn}$$\sim$0.009 \\
Reference                        & this work & Helfand et al. 2003   & Gaensler et al. 2002 \\
\enddata
\end{deluxetable}

As shown in Figure 2 (left), the pulsar has an associated extended
X-ray component aligned nearly north-south. In order to compare
its spatial characteristics with \chan's point-spread-function
(PSF), we performed a 2-D spatial fit to the data using the
GAUSS2D function in $\it{Sherpa}$ v2.2. First, images of the
source in the soft (0.5--1.15 keV), medium (1.15--2.15 keV) and
hard (2.15--10.0 keV) energy bands were created. The number of
counts in these energy images was 25, 170, and 90, respectively.
Corresponding normalized PSF images at an off-axis angle of
1$\farcm$28 and energies characteristic of the source's energy
histogram (0.85 keV, 1.7 keV and 3.0 keV) were made and used as
convolution kernels when fitting. From this fit, the low energy
image yielded a FWHM value fully consistent with the PSF
($\sim$0$\farcs$8), while the medium band image had a slightly
larger value ($\sim$0$\farcs$9). The hard band image was best
described by an elliptical gaussian function with FWHM of
$\sim$0$\farcs$9$\times$1$\farcs$2, confirming the extended nature
of the source. We rule out trailing in the S3 chip as an origin
for this feature since the read-out direction is at an angle of
$\sim$35$\degr$ measured counter-clockwise from the almost
north-south direction of the extended emission.

In order to further characterize the morphology of this extended
emission, we smoothed the above energy images using a Gaussian
with $\sigma$ = 0$\farcs$5 and then normalized them. The
contamination by the surrounding SNR in these images is small
($<$7\% of total). We then subtracted the normalized soft image
(consistent with a point source) from the normalized hard image.
Figure 3 shows the resulting image, which reveals structures that
appear to be consistent with torus- and jet-like features
surrounding the pulsar. Such features have been observed around
young rotation-powered pulsars (e.g. Lu et al. 2002) and are
believed to be associated with the deposition of the pulsar's
energy into its surroundings.

We estimate a total of 41 background-subtracted counts from an
annulus centered at the X-ray coordinates with radius
1$\farcs$2--3$\farcs$4 (see Fig. 3, right). (The inner radius was
chosen to be $\sim$1.5 times the PSF so the contamination from the
point source would be less than 15\%. The outer radius includes
the emission detected from the extended component.) This
corresponds to a significance of detection $\sigma$$\sim$5.5 and
contributes $\sim$14\% to the total count rate from the point
source and extended component. From the 41 counts quoted above, 14
are present in the medium energy band (1.15--2.15 keV) and 25 in
the high energy band (2.15--10.0 keV). A spectral analysis of the
extended emission is not possible with the available number of
counts and has to await a deeper exposure. We note however that,
unlike the point source, the extended emission is not detected in
the soft band (see Fig.~3, left). This leads us to conclude that
the extended emission appears to be harder than the point source
(assuming both have the same column density).

\subsection{SNR \g292}
We defer a detailed study of the emission from the SNR and the
newly identified point sources to a follow-up paper. Here, we
briefly summarize the results of our spectral fits targeted to
derive the column density needed to constrain the distance to the
system (see \S3.1). A spectrum of the SNR's interior was extracted
from the S3 chip. The point sources were excluded and source-free
regions were used as background.
A single component non-equilibrium ionization (NEI) model
 provided an acceptable fit.
The spectral parameters derived from our best fit VPSHOCK model
are: $N_{H}$=(5.8$\pm$1.0)$\times$10$^{21}$~cm$^{-2}$,
$kT$=36$\pm$10~keV and
$n_{e}t$=(5.1$\pm$1.0)$\times$10$^{9}$~cm$^{-3}$~s, with a
$\chi^{2}_{\nu}$(dof)=1.04 (278). A two-component NEI thermal and
non-thermal model provided an equally acceptable fit, with the
non-thermal model well described by a power law with a hard photon
index ($\Gamma$$\lesssim$1). Regardless of the model, we found an
$N_{H}$ value similar to the one quoted above. This value is
consistent with, but better constrained than, the one derived for
the pulsar and its PWN (see Table~1). It is also intermediate
between the $\it{ASCA}$ values derived for the eastern and western
sides of the remnant, $\sim$14$\times$10$^{21}$~cm$^{-2}$ and
$\sim$2$\times$10$^{21}$~cm$^{-2}$, respectively (Pivovaroff et
al. 2001).

\section{Discussion}

\subsection{Distance}
The extinction per unit distance in the direction of the system
can be estimated to be $E_{B-V}/D$ $\sim$ 0.2 mag~ kpc$^{-1}$
(Lucke 1978). We can then use the relation $N_{H}/E_{B-V}$ =
5.55$\times$10$^{21}$~cm$^{-2}$~mag$^{-1}$ to estimate the
distance. From our $N_{H}$ values from Table 1 and \S2.2, we
obtain a distance of 5.4--12.6 kpc for the pulsar and 4.0--6.3 kpc
for the remnant. A distance of 4--8 kpc was then adopted for the
system. The upper limit is determined from the location of the
source with respect to the Carina spiral arm (Camilo et al. 2000).

\subsection{PSR J1119--6127 and its PWN}
The analysis described in \S2.1 allowed us to identify the X-ray
counterpart of PSR J1119--6127 and its PWN. Using the spectral
model outlined in Table 1, we derive a 0.5--10.0 keV X-ray
luminosity for the point source and PWN of $L_{X}$ =
(5.5$^{+10}_{-3.3}$)$\times$10$^{32}$D$^{2}_{6}$~ergs~s$^{-1}$,
where D$_{6}$ is the distance in units of 6 kpc. The conversion
efficiency of $\dot{E}$ into $L_{X}$ is then $\epsilon_{psr+pwn}$
=
($L_{X}$/$\dot{E}$)$\sim$(2.4$^{+4.5}_{-1.5}$)$\times$10$^{-4}$D$^{2}_{6}$.
This value is somewhat on the low end of the efficiencies
exhibited by other pulsars associated with SNRs (e.g. Safi-Harb
2002).

In Table 2, we summarize the properties of J1119--6127 and two
other pulsars with similar spin properties. PSR \j1846\ lies
within $1'$ of the center of SNR Kes 75 and PSR \b1509\ lies close
to the center of SNR G320.4--1.2. All three pulsars spin slowly
($P$$>$100 ms) and have large inferred magnetic fields
($B$$>$1$\times$10$^{13}$ G) compared to other young, Crab-like
pulsars. However, their observed X-ray properties are different
from those of J1119--6127. While PSR \j1846\ exhibits the highest
X-ray efficiency (even when compared to all other Crab-like
pulsars, it has one of the highest $\epsilon$ values), J1119--6127
exhibits the lowest value ($\epsilon_{psr+pwn}$ $\lesssim$ 0.001,
using the upper limits on the luminosity and distance).
Furthermore, both \j1846\ and \b1509\ exhibit very similar
spectral properties, with the measured photon indices of the X-ray
pulsars being flatter than those of their associated PWNe. On the
other hand, as noted above, our analysis of PSR J1119--6127
suggests the opposite trend, with the photon index of the point
source being steeper than that of the extended feature. (We here
caution the reader that this conclusion is based on the small
number of counts available). Therefore, it seems that their
peculiar spin properties and high magnetic fields cannot account
in an obvious way for the differences in their X-ray properties.

\subsection{Relation to \axj}
The nature of the X-ray source \axj, detected with $\it{ASCA}$'s
Gas Imaging Spectrometer (GIS), is also of interest. This source
does not have a \chan\ point-source counterpart (see Fig.~4). The
coordinates of \axj\ were reported to be $\alpha$$_{J2000}$ =
11$^{h}$19$^{m}$03$\fs$4 and $\delta$$_{J2000}$ =
--61$\degr$$28'$$30''$, with an error radius of $24''$ (Pivovaroff
et al. 2001). This represents an offset of $87''$ from PSR
J1119--6127. We have re-examined the $\it{ASCA}$ data and found a
similar offset. Using an aperture of 2$\farcm$5 centered on the
$\it{ASCA}$ source, a background-subtracted count rate of
$\sim$0.004 s$^{-1}$ in the 3.0--10.0 keV range was reported with
the GIS. The count rate obtained with \chan\ using the same
region, which includes emission from the pulsar, PWN and
surrounding SNR, is consistent with the above GIS count rate. We
then conclude that \axj\ represents the unresolved pulsar
counterpart contaminated by its surrounding emission. Finally, the
large offset between \axj\ and the pulsar can be attributed to a
combined effect of $\it{ASCA}$'s limited resolution and additional
positional errors (e.g. Ueda et al. 2000 and Gotthelf et al.
2000).

\section{Conclusions}
Thanks to the sensitivity and resolution offered by \chan, we have
detected the X-ray counterpart of PSR J1119--6127 associated with
\g292, and resolved a $3''$$\times$$6''$ extended diffuse emission
that represents the first evidence for its PWN. Additional deep
observations of this system are needed to: 1) better constrain the
pulsar parameters, 2) study in detail the morphology of the PWN,
3) determine its spectral properties independently from the point
source, and 4) address the nature of the hard X-ray emission from
the interior of \g292. A timing observation will also allow the
search for pulsations from the X-ray source. Our observation
illustrates the need for $\it{Chandra}$ to unveil small PWNe that
could be associated with high-magnetic field radio pulsars and to
resolve their powering engines. In addition, our study suggests
that the X-ray properties of PSR J1119--6127 are different from
those of other pulsars with similar spin and magnetic field
properties, and therefore this class of pulsars merits further
study.

\acknowledgments

The authors thank the referee, Fernando Camilo, for his invaluable
comments that improved the paper. This research made use of NASA's
Astrophysics Data System (ADS) and the High Energy Astrophysics
Science Archive Research Center (HEASARC) operated by NASA's
Goddard Space Flight Center. The authors acknowledge support by
the Natural Sciences and Engineering Research Council (NSERC) of
Canada.




\onecolumn



\begin{figure}
\begin{center}
\centerline
{\includegraphics[scale=1]{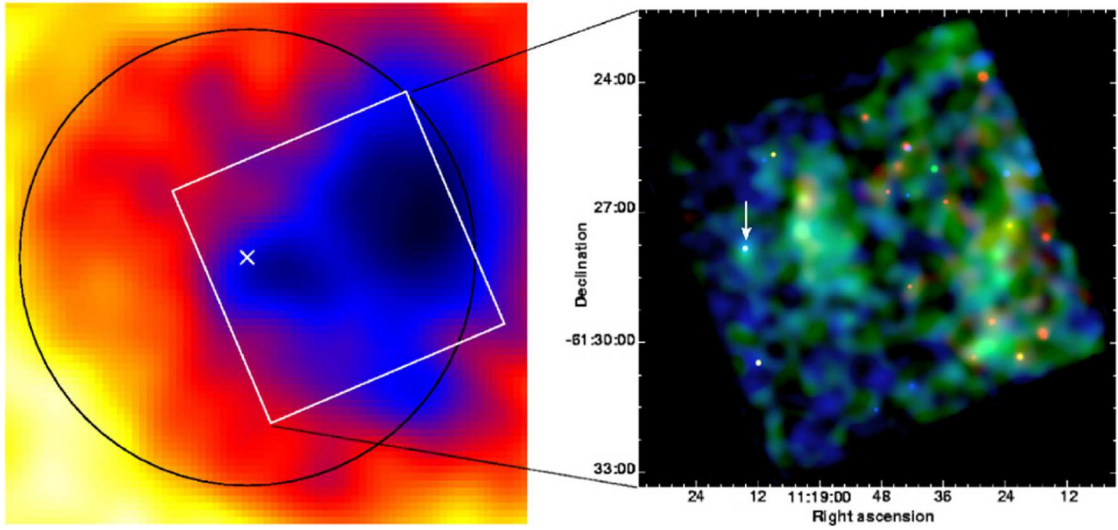}}
\caption{\label{1}$Left$:
2.0--10.0 keV $\it{ASCA}$ image of SNR \g292. The superimposed
regions mark the locations of \chan's S3 chip (white box), the
radio coordinates of PSR J1119--6127 (white cross) and the radio
boundary of \g292\ (black circle, $15'$ diameter). $Right$: \chan\
ACIS-S3 image of the interior of the remnant. Individual images in
the soft (0.5-1.15 keV, $\it{red}$), medium (1.15-2.3 keV,
$\it{green}$), and hard (2.3-10.0 keV, $\it{blue}$) bands were
combined. Resolution ranges from $1''$-$10''$ and black regions
represent non-significant detection. The arrow marks the location
of the detected counterpart of PSR J1119--6127.}
\end{center}
\end{figure}


\begin{figure}
\begin{center}
\centerline {\includegraphics[scale=1]{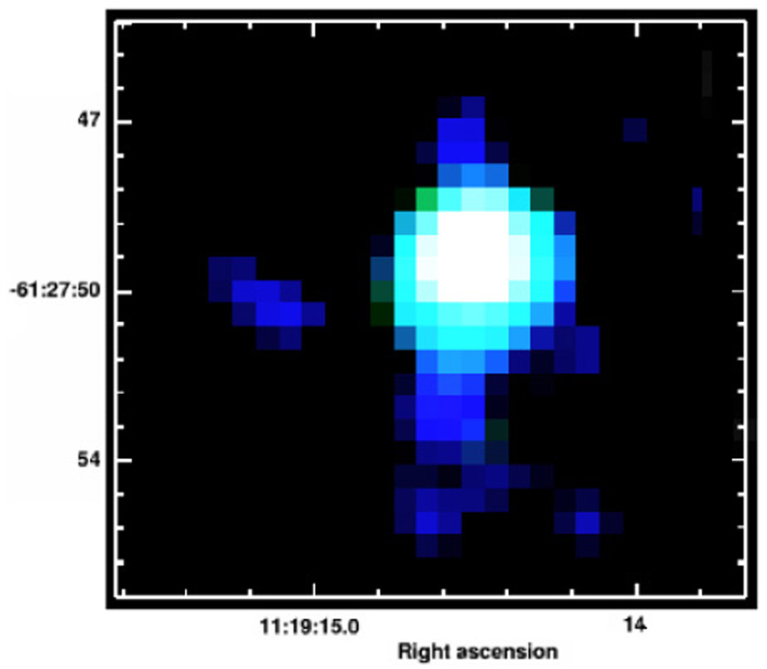}
\includegraphics[scale=1.1]{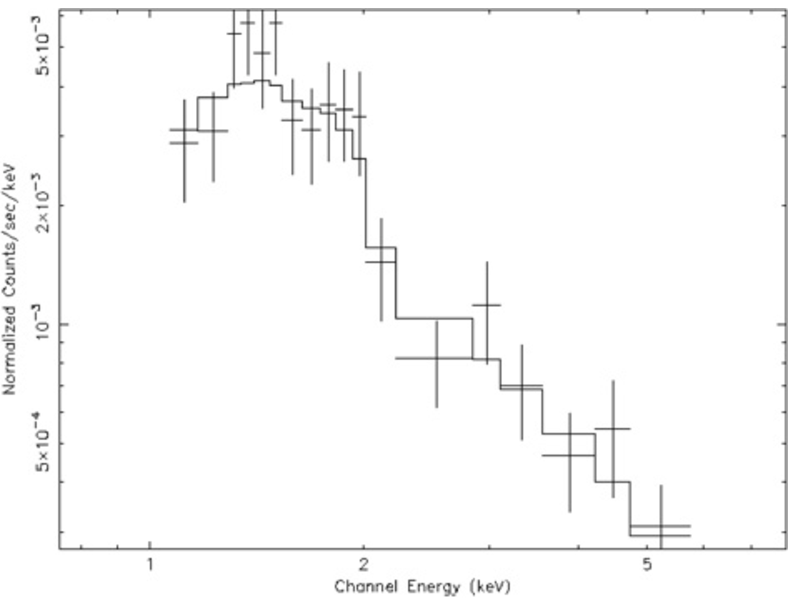}}
\caption{\label{2} $Left$: The X-ray counterpart of PSR
J1119--6127. This `truecolor' image combines the soft (0.5--1.15
keV, $red$), medium (1.15--2.15 keV, $green$), and hard (2.15--10
keV, $blue$) bands. The images were smoothed with a Gaussian with
$\sigma$=0$\farcs$5. $Right$: Spectrum and best fit power-law
model (see Table 1). A minimum significance of 3$\sigma$ was
chosen in rebinning the data for display.}
\end{center}
\end{figure}


\begin{figure}
\begin{center}
\centerline {\includegraphics[scale=1.2]{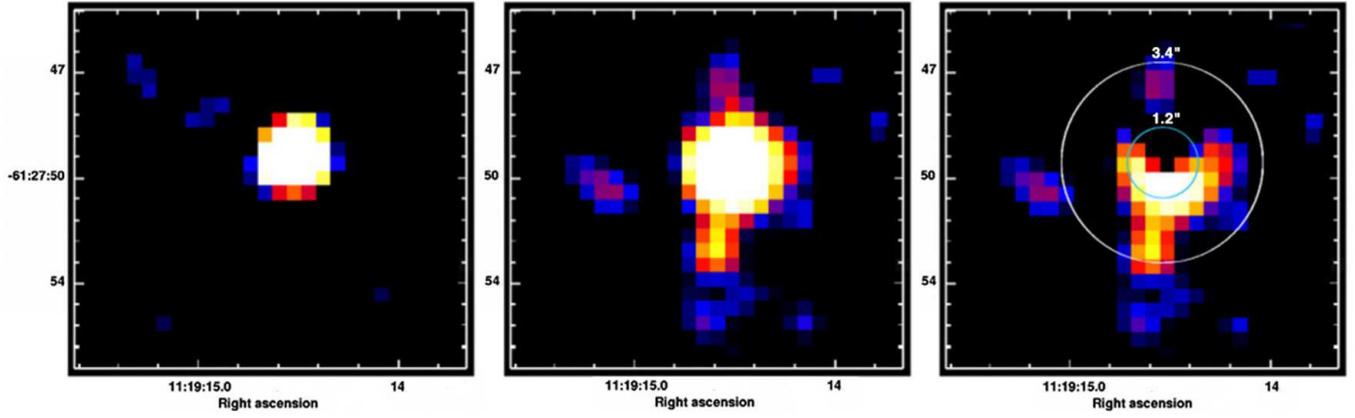}}
\caption{\label{3} $Left$ and $Center$: Normalized soft (0.5--1.15
keV, left) and hard (2.15--10.0 keV, center) energy images of PSR
J1119--6127 smoothed with a Gaussian with $\sigma$=0$\farcs$5.
$Right$: subtracting the soft (left) from the hard (center) band
image makes the underlying structure surrounding the pulsar
visible. A logarithmic display scale was used. See \S2.1 for
details.}
\end{center}
\end{figure}


\begin{figure}
\begin{center}
\centerline {\includegraphics[scale=2]{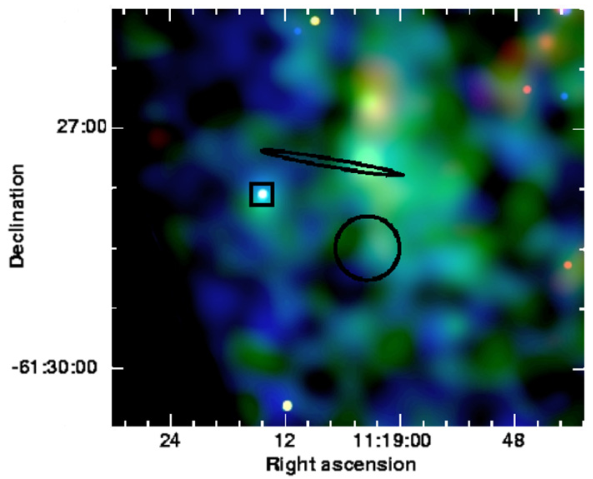}}
\caption{\label{4} 0.5--10.0 keV \chan\ image of the region around
the $\it{ASCA}$ source \axj. The small box marks the position of
PSR J1119--6127 (error much less than displayed size), the ellipse
shows a nearby IRAS source, J11169--6111 (error of
$53''$$\times$$3''$), and the circle shows the location of \axj\
(error radius of $24''$). See \S3.3 for details.}
\end{center}
\end{figure}


\begin{thebibliography}{}
\bibitem[]{375} Camilo, F. et al.  2000, \apj, 541, 367
\bibitem[]{376} Crawford, F. et al.  2001, \apj, 554, 152
\bibitem[]{377} Gaensler, B. M., Arons, J., Kaspi, V. M., Pivovaroff, M. J., Kawal, N. \& Tamura,
K. 2002, \apj, 569, 878
\bibitem[]{379} Gotthelf, E. V., Ueda, Y., Fujimoto, R., Kii, T., \& Yamaoka, K. 2000, \apj, 543, 417
\bibitem[]{378} Helfand, D. J., Collins, B. F. \& Gotthelf, E. V. 2003, \apj, 582, 783
\bibitem[]{380} Lu, F. J., Wang, Q. D., Aschenbach, B., Durouchoux, P. \& Song, L. M. 2002, \apj, 568, 49
\bibitem[]{381} Lucke, P. B. 1978, A\&A, 64, 367
\bibitem[]{386} Pivovaroff, M. J., Kaspi, V. M., Camilo, F., Gaensler, B. M., \& Crawford, F. 2001, \apj,
554, 161
\bibitem[]{388} Townsley, L. K., Broos, P. S., Garmire, G. P., \& Nousek, J. A. 2000, \apj, 534, L139
\bibitem[]{389}Safi-Harb, S. 2002, in  Proceedings of the 4th Microquasar Workshop, eds. Ph Durouchoux,
Y. Fuchs and J. Rodriguez, published by the Center for Space Physics: Kolkata (in press)
\bibitem[]{391} Ueda, Y. et al. 2000, AdSpR, 25, 839
\end{thebibliography}
\end{document}